\documentstyle[prl,aps,amsfonts,epsf]{revtex}
\begin{document}
%\draft
\title{Diffusional growth of wetting droplets}
\author{R.~Burghaus}
\address{Institut f{\"u}r Theoretische Physik IV,
  Heinrich-Heine-Universit{\"a}t D{\"u}sseldorf, Universit{\"a}tsstr.~1,
  D-40225 D{\"u}sseldorf, Germany}
\maketitle

\begin{abstract}
  The diffusional growth of wetting droplets on the boundary wall of a
  semi-infinite system is considered in different regions of a
  first-order wetting phase diagram. In a quasistationary
  approximation of the concentration field, a general growth equation
  is established on the basis of a generalized Gibbs-Thomson relation
  which includes the van der Waals interaction between the droplet and
  the wall. Asymptotic scaling solutions of these equations are
  found in the partial-, complete- and pre-wetting regimes.
\end{abstract}


The physics of wetting phenomena has attracted much interest in recent
years, both from experimental \cite{Bonn,DupontRoc,Taborek} and from
theoretical \cite{Cahn,Ebner,deGennes,Lenz,Dietrich,Lipowsky} points of
view.  Whereas initially static properties dominated the discussion,
the interest has shifted more recently to the dynamics of wetting
\cite{deGennes,Leger,Anderson,Oron,Deegan,Troian,Wiltzius,Lipowsky2,Herminghaus}.
In many experimental situations the formation of a wetting layer
starts with the nucleation of droplets on the boundary wall of the
system. The central question therefore is the temporal evolution of
the droplet profile.

There are essentially two different types of dynamic behavior of a
liquid surface droplet. The first is a spreading process which
e.g. dominates, if a droplet of a non-volatile liquid is overheated
from below to above a wetting transition point. Such processes are
driven by hydrodynamic modes of the liquid, and they have extensively
been discussed in the literature
\cite{deGennes,Leger,Anderson,Oron}. The second mechanism is the phase
transformation (condensation or evaporation) between the liquid and
the vapor phase of the droplet. This is driven by particle diffusion
in the vapor, and e.g. is the dominating process in the growth of
supercritical droplets in a metastable situation.

Whereas the diffusional growth of a homogeneous wetting layer has been
discussed in the literature \cite{Lipowsky2}, this seems not to be the
case for surface droplets. The present paper deals with the
diffusional growth of a supercritical droplet from a super-saturated
vapor. This process is accompanied by the creation of latent heat, and
it will be assumed that heat transport as well as other hydrodynamic
modes are fast compared to the diffusion. As a consequence the droplet
is isothermal and always has a shape which minimizes its free energy
at a given volume. The time dependence of this shape is the main
object of interest in this paper.

\begin{figure}[h]
\begin{center}
\leavevmode
\vbox{
\epsfxsize=4in
\epsffile{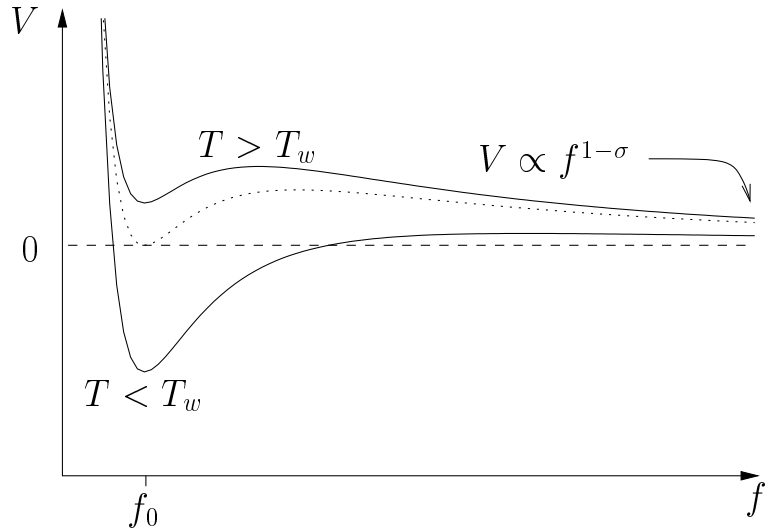}}
\end{center}
 \caption{Sketch of an effective interface potential that shows a
    first order wetting transition as $T$ is raised from $T<T_w$ to
    $T>T_w$.}
  \label{fig:V}
\end{figure}

The excess free energy of a wetting film of local thickness $f(x)$
on a planar boundary wall of a semi-infinite system can be
written in the form \cite{Brezin,Lipowsky3}
\begin{equation}
  \label{InterfaceModell}
  {\cal H}_h[f] = \int{d^2x\,\left[{\gamma\over 2}\left(\nabla
  f\right)^2 + V(f) - h\,f\right]},
\end{equation}
where $\gamma$ is the interface stiffness, $h$ is the difference of
the chemical potential from that of the saturated vapor and $V(f)$ is
an effective interface potential. The field $h$ can be expressed by
the difference between the vapor concentration $c$ and its value $c_0$
at saturation, so that in linear order
\begin{equation}
  \label{c_h}
  c = c_0 \left( 1 + \Gamma h \right).
\end{equation}
The form of the potential $V(f)$ corresponding to a first-order
wetting transition is sketched in Fig.~\ref{fig:V}. There, for
temperatures $T$ less than the wetting temperature $T_w$, the global
minimum of $V(f)$ is at $f=f_0$, whereas for $T>T_w$ this minimum
becomes metastable in favor of the global minimum at diverging film
thickness.  For $f\to\infty$ we assume $V\propto f^{1-\sigma}$, where
$\sigma=3$ for nonretarded and $\sigma=4$ for retarded van der Waals
interactions \cite{deGennes}.

\begin{figure}[h]
\begin{center}
\leavevmode
\vbox{
\epsfxsize=5in
\epsffile{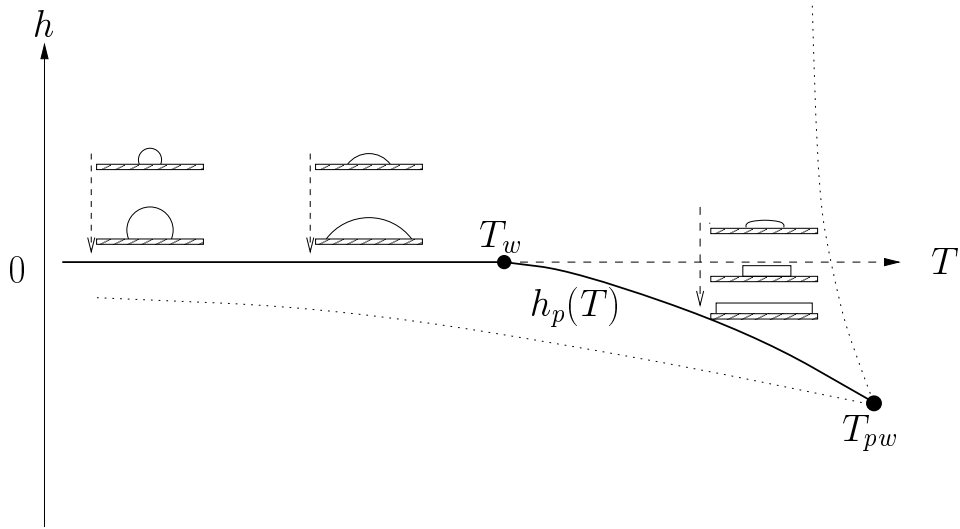}}
\vspace{10mm}
\end{center}
  \caption{Wetting phase diagram: The first order line of
    wetting transitions consisting of the partial wetting line $h=0,
    T<T_w$ and the prewetting line $h=h_p(T)$ is marked by a solid
    line. The different types of growing wetting droplets along
    isotherms are shown in the different regions of the phase diagram.
    The dotted lines refer to the surface spinodals which enclose the
    nucleation regime.}
  \label{fig:PhaseDiag}
\end{figure}

Homogeneous (i.e.~$f(x)=const$) minima of the excess free energy
${\cal H}_h[f]$ (i.e. the {\sl global} minima of $V(f)-hf$) determine
the phase diagram, shown in Fig.~\ref{fig:PhaseDiag}. In the region
$h>0$, where the liquid bulk phase is stable, a film of infinite
thickness forms on the wall in thermal equilibrium. On the line $h=0$,
which means bulk coexistence of the liquid and vapor phases, the
first-order wetting transition occurs at $T=T_w$, where $f=f_0$ for
$h=-0$, $T<T_w$ (partial wetting), and $f=\infty$ for $T>T_w$
(complete wetting).  From the transition point a prewetting line
$h_p(T)$ extends into the region $h<0$ where the vapor phase is stable
in the bulk. This line separates a region (below $h_p(T)$) where the
wall is covered by a thin film from a region ($h>h_p(T)$) where the
wall is covered by a thick film. The jump in film thickness along the
prewetting line vanishes at the prewetting critical point $T_{pw}$.
The partial wetting line $h=0$, $T<T_w$ and the prewetting line
$h_p(T)$ together form a first order line concerning the wetting
properties of the system \cite{Bausch4}.

If the system is quenched from below to above the first order line
(for example by increasing the pressure), the phase transition is
initialized by the formation of critical droplets on the wall
(provided, one stays within the surface spinodal lines, shown in
Fig.~\ref{fig:PhaseDiag}). The shape of these droplets is
qualitatively different in different regions of the phase diagram
\cite{Bausch2,Bausch,Bausch3}. Axisymmetric profiles $f(r)$ can be
calculated via the saddle point equation
\begin{equation}
  \label{saddelpoint}
  \delta{\cal H}_h\,/\,\delta f(r) = 0
\end{equation}
with the natural boundary conditions of a droplet profile
\begin{equation}
  \label{bc}
  f^\prime(0)=0 , \quad \lim_{r\to\infty}f(r)=f_0.
\end{equation}
As illustrated in Fig.~\ref{fig:PhaseDiag}, this leads to spherical
(in the squared gradient approximation of Eq.~(\ref{InterfaceModell})
parabolic) caps in the partial wetting regime, to flat cylindrical
droplets (pancakes) in the prewetting regime \cite{deGennes} and to
ellipsoid-like droplets in the complete wetting regime \cite{Bausch2}.
The saddle point $\delta{\cal H}_h\,/\,\delta f(r) = 0$ has an unstable
growth mode but the volume preserving shape fluctuations are stable
\cite{Bausch,Bausch3}.

Assuming that the volume growth of the droplet is slow, the diffusion
in the surrounding concentration field $c$ becomes quasistationary and
can be approximated by the Laplace equation
\begin{equation}
  \label{Diffusion}
  D\;\Delta c = 0,
\end{equation}
where $D$ is the diffusion constant. At far distances from the droplet
the concentration field is given by the system concentration
$c_\infty(t)$ which is time-dependent in a supersaturated system
($h>0$) because of the phase-separation process in the metastable bulk
phase. The normal derivative of the concentration field on the
boundary wall of the system vanishes because there is no diffusion
flux into the wall, i.e. the Neumann boundary condition
\begin{equation}
  \label{Neumann}
  D\;\left.\partial_\perp c\right|_{\text{wall}} = 0
\end{equation}
has to be fulfilled.

To obtain a well defined diffusion problem the boundary condition on
the surface as well as the actual shape
of the supercritical droplet need to be specified. Motivated by
the slow diffusional growth of the droplet the concentration field
close to the droplet surface is assumed to be in local thermal
equilibrium.  Therefore the local chemical potential $h(x)$ at the
droplet surface is given by $h(x)={\delta{\mathcal H}_0}\,/\,{\delta
  f(x)}$ which due to (\ref{c_h}) corresponds to a concentration
\begin{equation}
  \label{Gibbs-Thomson}
  c_s(x) = c_0\left(1 + \Gamma{\delta{\mathcal H}_0 \over \delta
  f(x)}\right),
\end{equation}
which can be denoted as a generalized Gibbs-Thomson relation for
wetting droplets. The expression $\delta {\cal H}_0\, / \,\delta f$
consists of a term $\gamma$ times the local curvature $K$ of the
droplet interface plus an interaction term $\partial V \,/\,\partial
f$. Neglect of the interaction term reduces (\ref{Gibbs-Thomson})
to the classical Gibbs-Thomson relation $c_s=c_0(1+\Lambda K)$ with
the capillary length $\Lambda\equiv\Gamma\gamma$. It identifies the
concentration at a curved interface as the concentration $c_0$ for a
flat interface modified by a linear curvature correction.

The assumption of fast hydrodynamic modes (compared with the
diffusional growth) implies that the shape of a growing droplet can be
calculated by minimizing its free energy under the constraint of a
fixed droplet volume $\Omega(t)$. Technically, this variational
calculation leads again to Eq.~(\ref{saddelpoint}) with
the boundary conditions~(\ref{bc}) but now with $h$ in
(\ref{saddelpoint}) replaced by a function $h_{\Omega(t)}$ which
includes a Lagrange parameter corresponding to $\Omega(t)$. Consequently
the growing supercritical droplet always looks like a critical droplet
at a different time-dependent chemical potential. With increasing
volume $\Omega(t)$ the corresponding field $h_{\Omega(t)}$ approaches
the first order line, where eventually the volume of the droplet
diverges.  In this sense wetting droplets grow along isotherms towards
the first order line, as illustrated in Fig.~\ref{fig:PhaseDiag}.

The saddle point equation (\ref{saddelpoint}) with the fixed volume
constraint is equivalent to $\delta{\cal H}_0\,/\,\delta
f=h_{\Omega(t)}$. Via Eq.~(\ref{Gibbs-Thomson}) this implies that the
Dirichlet boundary condition of the constrained equilibrium droplet is
given by
\begin{equation}
  \label{Dirichlet}
  c_s(t) = c_0\left(1 + \Gamma\;h_{\Omega(t)}\right)
\end{equation}
and therefore independent of $x$. Especially in the complete
wetting or prewetting case, where the droplets are not spherical, one
would expect a non trivial boundary condition having the classical
Gibbs-Thomson condition in mind. Additionally corrections due to the
potential $V(f)$, which determine the shape of the droplets in these
regions, have to be taken into account. Nevertheless both effects add
up in such a way that Eq.~(\ref{Gibbs-Thomson}) can be written as
Eq.~(\ref{Dirichlet}) for a droplet in a volume-constraint equilibrium
showing that $c_s$ is constant along the droplets surface!

Now, the $x$-independent Dirichlet boundary condition (\ref{Dirichlet})
allows to use an electrostatic analogy to solve the quasistationary
diffusion problem (\ref{Diffusion})--(\ref{Gibbs-Thomson}) for the
growing droplet \cite{Burghaus2}. To fulfill the Neumann condition
(\ref{Neumann}), the system (including the droplet) is mirrored at the
boundary wall of the system.  Then the field $4\pi D c$ is identified
with an electric potential which also obeys the Laplace equation. The
normal derivative of the field, i.e. the diffusion flux density on the
droplet surface field, corresponds to the charge density of a
conductor with the shape of the droplet including its mirror image.
Consequently, the total volume growth of the droplet corresponds to
the total charge, which is given by the capacity $C$ of the conductor
times the potential difference between the surface and infinity. This
ultimately leads to the droplet growth equation
\begin{equation}
  \label{growth}
  \dot\Omega = 4 \pi D C(t)\left[c_\infty(t) - c_s(t)\right] =
  4 \pi D \Gamma C(t)\left[h(t)-h_{\Omega(t)}\right]
\end{equation}
where $C$ depends on the droplets profile and therefore is implicitly
time-dependent \cite{Diss}. The difference
$\left[h(t)-h_{\Omega(t)}\right]$ may be interpreted as the
supersaturation of the system with respect to the droplet.

Eq.~(\ref{growth}) together with Eqs.~(\ref{InterfaceModell}),
(\ref{saddelpoint}) and (\ref{bc}) allow to determine selfconsistently
the growing droplet profile if $h(t)$ is known. For a
given volume $\Omega$, the droplet profile can be calculated from by
Eqs.~(\ref{InterfaceModell}), (\ref{saddelpoint}) and (\ref{bc}) with
a conveniently chosen Lagrange multiplier $h_\Omega$. Then the
capacity $C$ of the conductor represented by the droplet plus its
mirror image is calculated.  Insertion of $C$, $h_\Omega$ and the
chemical potential $h$ into Eq.~(\ref{growth}) yields the droplet
growth rate $\dot\Omega(t)$ and integration of (\ref{growth})
eventually determines $\Omega(t)$.

In practice, for large droplets, the calculation can be facilitated by
the use of scaling properties of critical droplets close to the
first-order transition line \cite{Bausch2,Bausch}. At temperatures
$T>T_w$ the wetting droplets on the wall nucleate either as
ellipsoid-like droplets at $h\ge 0$ (complete wetting) or as pancake
like droplets at $h_p(T)<h<0$ (prewetting). In both cases the droplets
grow along an isotherm towards the prewetting line
($h_{\Omega(t\to\infty)}\to h_p(T)$). This means that they eventually
become pancake-like droplets with a constant height but diverging
radius $R(t)$ so that the capacity of large droplets is given by
the capacity of a flat disk $C(t)\propto R(t)$. In the case where the
initial quench leads to a supersaturated bulk system ($h>0$) the
volume will phase separate until it reaches $h=0$ whereas for initial
values $h<0$ the vapor bulk phase is stable and $h$ remains constant
in time. In either situation the difference $[h-h_\Omega]$ approaches
a non-vanishing constant so that Eq.~(\ref{growth}) yields $\dot\Omega
\propto \Omega^{1/2}$ or $\Omega \propto t^2$ which implies
\begin{equation}
  R \propto t,
\end{equation}
and only is determined by the time dependent capacity, i.e. the
increasing diffusive coupling to the environment.

Wetting droplets at $T=T_w$ in a supersaturated system ($h>0$) also
are not spherical. Their radius $R$ scales as $R\propto
{h_\Omega}^{-(\sigma +1)\,/\,2\sigma}$, their central height $F$ as
$F\propto {h_\Omega}^{-1/\sigma}$, and consequently their volume as
$\Omega \propto {h_\Omega}^{-(\sigma+2)/\sigma}$ \cite{Bausch2}.
Therefore, the profile of a growing wetting droplet becomes flatter
and approaches a disk with capacity $C\propto R \propto
{h_\Omega}^{-(\sigma + 1)\,/\,2\sigma}$. In a supersaturated system
(i.e. $h>0$) there are not only wetting droplets on the wall, but also
droplets in the bulk.  The set of growing bulk droplets reduces the
supersaturation in a Lifshitz-Slyozov-Wagner type way as $h\propto
t^{-1/3}$ \cite{LifSly,Wagner,Burghaus}. With this input the wetting
droplet growth equation (\ref{growth}) can be written as
\begin{equation}
  \dot h_\Omega\,{h_\Omega}^{-{2(\sigma+1)\over \sigma}} \propto
  {h_\Omega}^{-{\sigma + 1 \over 2\sigma}} \left[A\;t^{-1/3} - h_\Omega\right]
\end{equation}
for large droplets. This leads to the asymptotic growth law
$h_\Omega \propto t^{-{4 \sigma \over 3(\sigma+3)}}$ which due to the
above scaling properties for $R$ and $F$ implies
\begin{equation}
  R \propto t^{2(\sigma +1) \over 3 (\sigma +3)}, \qquad F \propto
  t^{4 \over 3(\sigma +3)}.
\end{equation}

Finally, in the partial wetting regime $T<T_w, h>0$, the wetting
droplets are spherical caps, and therefore their growth properties are
similar to those of bulk droplets, i.e.
\begin{equation}
	R \propto t^{1/3}
\end{equation}

 The evaluation of the difference $[h-h_\Omega]$ in Eq.~(\ref{growth})
can only be done in a theory where diffusional interactions between
surface and bulk droplets are taken into account \cite{Burghaus2}. At
late stages partial wetting droplets in systems with a temperature
corresponding to a contact angle $\Theta>\pi/2$ will shrink because
$[h-h_\Omega]$ turns negative for each droplet, whereas droplets at a
temperature with $\Theta<\pi/2$ will grow.

One of the basic ingredients of the present calculation is the Neumann
boundary condition (\ref{Neumann}). It derives from the fact that
there is no diffusion flux through the wall. Even if the wall locally
is in a non-wet state it always is covered by a film of microscopic
thickness $f_0$. If somewhere in such a region Eq.~(\ref{Neumann})
were not valid, the film would thicken there and the interface would
run out of the microscopic minimum of the interface potential shown in
Fig.~\ref{fig:V}. Due to the generalized Gibbs-Thomson
relation~(\ref{Gibbs-Thomson}), which includes a term $\partial
V\,/\,\partial f$, the local concentration on top of the surface would
then increase and the film would evaporize until it again reaches the
former height $f_0$. Thus, up to fluctuations, Eq.~(\ref{Neumann})
will be valid.

The calculation of the supercritical droplet shape is based on the
assumption of fast hydrodynamic modes compared to the droplets
diffusional growth. This assumption may become questionable, if a
prewetting droplet becomes very large. However at this very late
stage the coalescence of different droplets will be dominant anyway.

I like to thank R.~Bausch and R.~Blossey for stimulating discussions
and the Deutsche Forschungsgemeinschaft via SFB~237 ``Unordnung und
gro{\ss}e Fluktuationen'' and ``Benetzung und Strukturbildung an
Grenzfl{\"a}chen'' as well as the EU via FMRX-CT~98-0171 ``Foam Stability
and Wetting Transitions'' for financial support.

\newpage

% Abbildungen

\vspace{20mm}

\end{document}